\documentclass[sigconf]{acmart}

\usepackage[switch]{lineno}
\newcommand{\toolname}{\textsc{Multi\textsuperscript{2}Fixer}\xspace}

\newcommand{\buganalyzer}{\textsc{BugAnalyzer}\xspace}
\newcommand{\coordinator}{\textsc{Coordinator}\xspace}
\newcommand{\proposer}{\textsc{Proposer}\xspace}

\usepackage{tcolorbox}
\tcbuselibrary{breakable}
\usepackage{pifont}
\usepackage{enumerate}
\usepackage{ragged2e}
\usepackage{makecell}
\usepackage{stfloats}
\usepackage[vlined, ruled]{algorithm2e}
\usepackage{algorithmic}
\usepackage{graphicx}
\usepackage{textcomp}
\usepackage{multirow}
\usepackage{subfigure}
\usepackage{diagbox}
\usepackage{listings}
\usepackage{float}
\usepackage{url}
\usepackage{hyperref}

\definecolor{deepblue}{rgb}{0,0,0.5}
\definecolor{deepgreen}{rgb}{0,0.5,0}
\definecolor{deepred}{rgb}{0.6,0,0}
\definecolor{darkorange}{RGB}{255,140,0}
\definecolor{lightgray}{rgb}{0.93,0.93,0.93}
\definecolor{deepgray}{rgb}{0.25,0.25,0.25}
\definecolor{diffgreen}{RGB}{218,245,225}
\definecolor{diffred}{RGB}{255,224,224}
\definecolor{diffblue}{RGB}{222,235,255}
\newcommand{\diffadd}[1]{\colorbox{diffgreen}{\makebox[\dimexpr\linewidth-2\fboxsep\relax][l]{\strut\textcolor{deepgreen}{#1}}}}
\newcommand{\diffdel}[1]{\colorbox{diffred}{\makebox[\dimexpr\linewidth-2\fboxsep\relax][l]{\strut\textcolor{deepred}{#1}}}}
\newcommand{\diffctx}[1]{\colorbox{white}{\makebox[\dimexpr\linewidth-2\fboxsep\relax][l]{\strut\textcolor{black}{#1}}}}
\newcommand{\diffhunk}[2]{\colorbox{diffblue}{\makebox[\dimexpr\linewidth-2\fboxsep\relax][l]{\strut\textcolor{deepblue}{\textbf{#1}}\hfill\textcolor{deepgreen}{#2}}}}

\hypersetup{
    colorlinks=true,
    linkcolor=blue,
    filecolor=blue,      
    urlcolor=blue,
    citecolor=blue,
}

\lstdefinelanguage{Java}{
	basicstyle=\small\ttfamily,
	numberstyle=\color{deepgray},
	stepnumber=1,
	numbersep=8pt,
	showstringspaces=false,
	breaklines=true,
	frame=lines,
	backgroundcolor=\color{lightgray},
	commentstyle=\color{deepgreen},
	keywordstyle=\color{deepblue},
	stringstyle=\color{deepred},
	tabsize=4,
	captionpos=b,
	morekeywords={public, class, void, int, if, else, for, while, return, true, false},
	emph={String, System},
	emphstyle=\color{darkorange},
	alsoletter={.,;:[]()},
}
\usepackage{multirow}
\usepackage[utf8]{inputenc}
\usepackage{graphicx}
\usepackage{textcomp}
\usepackage{color,xcolor}
\usepackage{url}
\usepackage{graphicx}
\usepackage{subfigure}
\usepackage{stmaryrd}
\usepackage{verbatimbox}
\usepackage{enumerate}
\usepackage[shortlabels]{enumitem}
\usepackage{bm}
\usepackage[justification=centering]{caption}
\newtheorem{definition}{Definition}
\usepackage{makecell}
\usepackage{booktabs}

\newcommand{\find}[2]{
\begin{tcolorbox}[toprule=0mm,bottomrule=0mm,left=1pt,right=2pt,top=2pt,bottom=2pt,breakable]\em #1
\end{tcolorbox}
}

\usepackage{multirow}

\usepackage{enumitem}
\usepackage{colortbl}

\AtBeginDocument{  \providecommand\BibTeX{{    \normalfont B\kern-0.5em{\scshape i\kern-0.25em b}\kern-0.8em\TeX}}}
    
\setcopyright{acmcopyright}
\acmPrice{15.00}
\acmYear{2026}
\copyrightyear{2026}
\acmConference[ASE'26]{Proceedings of the 41st IEEE/ACM International Conference on Automated Software Engineering}{October 12–16, 2026}{Munich, Germany}

\begin{document}
\settopmatter{authorsperrow=4}

\title{\toolname{}: A \coordinator{}-\proposer{} Based Multi-Agent Framework For Fixing Multi-Hunk Bugs}

\author{Haichuan Hu}
\affiliation{  \institution{Nanjing University of Science and Technology}
  \city{Nanjing}
  \country{China}
}
\email{huhaichuan2024@gmail.com}

\author{Chunrong Fang}
\affiliation{  \institution{Nanjing University}
  \city{Nanjing}
  \country{China}
}
\email{fangchunrong@nju.edu.cn}

\author{Ye Shang}
\affiliation{  \institution{Nanjing University}
  \city{Nanjing}
  \country{China}
}
\email{yeshang@smail.nju.edu.cn}

\author{Jiawei Liu}
\affiliation{  \institution{Nanjing University}
  \city{Nanjing}
  \country{China}
}
\email{jw.liu@nju.edu.cn}

\author{Weifeng Sun}
\affiliation{  \institution{Singapore Management University}
  \city{Singapore}
  \country{Singapore}
}
\email{weifeng.sun@cqu.edu.cn}

\author{Guoqing Xie}
\affiliation{  \institution{Nanjing University}
  \city{Nanjing}
  \country{China}
}
\email{522025320181@smail.nju.edu.cn}

\author{Chenxing Zhong}
\affiliation{  \institution{Nanjing University of Science and Technology}
  \city{Nanjing}
  \country{China}
}
\email{chenxingzhong@njust.edu.cn}

\author{Quanjun Zhang}
\authornote{corresponding author.}
\affiliation{  \institution{Nanjing University of Science and Technology}
  \city{Nanjing}
  \country{China}
}
\email{quanjunzhang@njust.edu.cn}

\begin{abstract}
Automated Program Repair (APR) has benefited greatly from Large Language Models (LLMs), but existing LLM-based APR methods still struggle with multi-hunk bugs that require coordinated changes across multiple locations. These bugs demand repository-level context understanding, repair-order scheduling, and effective hunk-level patch generation and selection. To address these challenges, we propose \toolname{}, a novel \coordinator{}-\proposer{} based multi-agent framework for multi-hunk repair.
\toolname{} performs tool-augmented bug analysis, constructs fine-grained repair context, iteratively generates patches through a \coordinator{}-\proposer{} architecture, and applies two-stage patch refinement for syntactic and semantic correctness. {}{We evaluate \toolname{} on 835 bugs from Defects4J and three vulnerability benchmarks.} On Defects4J, \toolname{} fixes 326 bugs, including 62 multi-method and 27 multi-file bugs, and outperforms prior APR baselines in the reported comparisons with the same base model. Moreover, \toolname{} also fixes {}{46} multi-hunk bugs among {}{95} unique fixes. When combined with Claude-3.5-Sonnet, \toolname{} repairs 420 bugs, establishing a new state of the art on Defects4J. On VUL4J, \toolname{} repairs 24 real-world vulnerabilities, including 5 multi-hunk cases. {On the multi-hunk subsets of SEC-bench and PatchEval, \toolname{} fixes 11 and 19 vulnerabilities, respectively, outperforming all compared baselines under GPT-3.5.} These results demonstrate the effectiveness of \toolname{} for multi-hunk repair.

\end{abstract}



\maketitle

\section{Introduction}
The emergence of Large Language Models (LLMs) has driven advancements in software engineering~\cite{zhang2023survey,hou2023large}, demonstrating outstanding capabilities across various code-related tasks, including code generation~\cite{jiang2024survey,du2024evaluating}, test generation~\cite{zhang2025large,shang2025large,shang2026breaking,zhang2026reproagent}, and program repair~\cite{zhang2024systematic,zhang2023gamma,hu2025tsapr,11029663,hu2026evorepair}. 
Building on this, LLM-based Automated Program Repair (APR) leverages LLMs for fault localization, patch generation, and patch validation, enabling the automatic detection and repair of software defects and attracting growing attention in modern software engineering. However, most existing LLM-based APR approaches primarily focus on relatively simple bugs~\cite{xia2024automated,yin2024thinkrepair}, and typically treat the bug as a whole in a single-step repair process~\cite{xia2024automated}. 
Such settings, while effective for controlled evaluation, exhibit limitations when applied to real-world program bugs.

\begin{figure}[htbp]
  \centering
  \includegraphics[width=\linewidth]{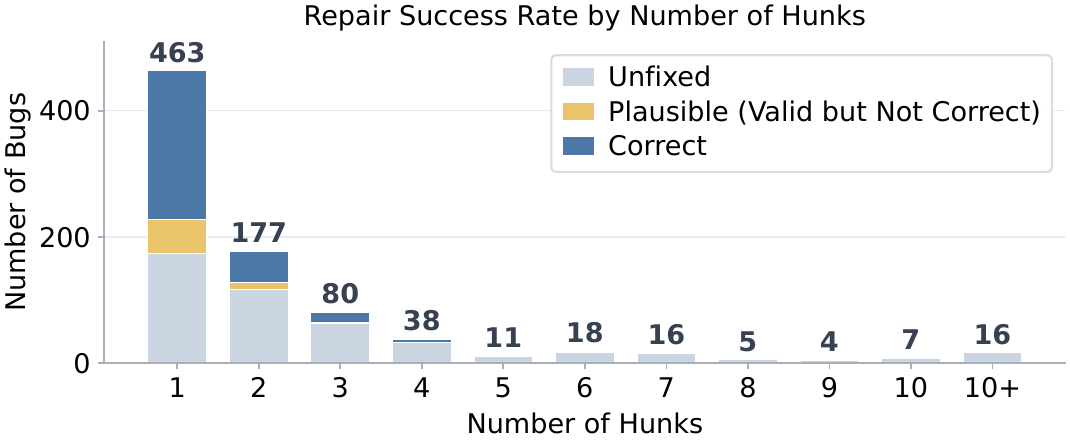}
  \caption{The relationship between repair success rate and the number of hunks.} 
  \label{fig:repair_success_rate_by_hunk_num}
\end{figure}

One major limitation is that a significant portion of real-world bugs involve complex, multi-hunk changes spread across different methods and files. One recent study~\cite{xie2025premm} analyzes the distribution of hunk counts and its impact on repair success rate across 835 bugs in Defects4J~\cite{just2014defects4j}. As shown in Figure~\ref{fig:repair_success_rate_by_hunk_num}, 
nearly 50\% of these bugs involve multiple hunks, and the repair success rate significantly decreases as the number of hunks increases. 
However, existing works such as ChatRepair~\cite{xia2024automated} and ThinkRepair~\cite{yin2024thinkrepair} focus solely on evaluating single-line, single-hunk, and single-method bugs, comprising 483 of the 835 bugs from Defects4J (i.e., 255 from Defects4J-v1.2 and 228 from Defects4J-v2.0). This means that the remaining 352 more complex bugs are excluded from their evaluations. Such selective evaluation substantially undermines the objectivity of assessing LLMs' true repair capabilities.

Another limitation is that existing methods treat multi-hunk bug repair as a single-step task, generating patches for all hunks within a single interaction. Studies~\cite{xie2025premm,ye2024iter} show that there are often dependencies among different hunks. By referring to how humans repair multi-hunk bugs, understanding and modifying hunks is a hierarchical and sequential process. Moreover, as the number of hunks increases, it becomes difficult for repair models to generate patches for all hunks correctly in a single attempt; if even one hunk is incorrect, the entire patch must be discarded.

These two limitations pose the following challenges for multi-hunk bug repair. 
(1) \textbf{Repository-level repair context.} 
Single-hunk bug repair typically requires only test failure information and buggy code as repair context, but fixing multi-hunk bugs requires repository-level global context. 
(2) \textbf{Repair order scheduling.} 
Due to semantic and logical dependencies among hunks, there is a natural repair order that must be followed. 
Identifying these inter-hunk relationships and properly scheduling the repair sequence is crucial. 
(3) \textbf{Hunk-level patch generation and selection.}
In the process of repairing multi-hunk bugs, we cannot use test cases to verify the correctness of individual hunk-level partial patches because the interactions between different hunks can lead to dependencies that obscure the effects of each individual patch, making isolated testing ineffective.

\textbf{This Paper.}
To address the above issues, we propose \toolname{}, a \coordinator{}-\proposer{} based multi-agent framework for fixing multi-hunk bugs. \toolname{} consists of four stages: (1) tool-augmented bug analysis with \buganalyzer{} to identify root causes and relevant code; (2) repair context construction by combining bug reports, buggy code, local context, test code, and failure reports; (3) iterative patch generation through a \coordinator{}-\proposer{} architecture, where the \coordinator{} schedules hunk repair and selects promising candidates from multiple \proposer{}s; and (4) two-stage patch refinement for syntactic and semantic correctness. The framework is motivated by the Generation-Recognition Asymmetry phenomenon~\cite{fan2024oracle}, suggesting that selecting a correct patch from candidates is easier than generating it directly.
We evaluate \toolname{} on {}{Defects4J~\cite{just2014defects4j} and three vulnerability benchmarks: VUL4J~\cite{bui2022vul4j}, SEC-bench~\cite{lee2025secbench}, and PatchEval~\cite{wei2025patcheval}}. 
Experimental results show that \toolname{} can fix 326 out of 835 bugs in Defects4J, including 62 multi-method and 27 multi-file bugs, outperforming prior baselines in our reported comparisons.
On VUL4J, \toolname{} fixes 24 vulnerabilities, including 5 multi-hunk vulnerabilities, {}{suggesting that \toolname{} can generalize to vulnerability repair}.
{On the SEC-bench and PatchEval multi-hunk subsets, \toolname{} further fixes 11 and 19 vulnerabilities, respectively, showing consistent advantages beyond Java vulnerability repair.}
To sum up, the main contributions of this paper are as follows:

\begin{itemize}[leftmargin=*]
    \item \textbf{New Mechanism.}
    For multi-hunk repair, we draw inspiration from the Generation-Recognition Asymmetry phenomenon and innovatively propose the \coordinator{}-\proposer{} repair mechanism.  We replace the traditional single-step paradigm by combining the \coordinator{}-\proposer{} mechanism with iterative hunk-level repair, thereby breaking through the model's capability boundaries in the multi-hunk repair scenario.

    \item \textbf{Well-Designed Technique.} Guided by the \coordinator{}-\proposer{} mechanism, we propose and implement a multi-agent repair framework, \toolname{}, which consists of three distinct agents: \buganalyzer{}, \coordinator{}, and \proposer{}. These agents collaborate to complete bug analysis and repairs. Notably, during patch generation, \toolname{} can autonomously schedule the repair order and revise existing hunk-level patches, thereby facilitating multi-hunk repairs.
    
    \item \textbf{Extensive Study.}
    {}{We evaluate \toolname{} on Defects4J and three vulnerability benchmarks, comparing its performance against state-of-the-art baselines. Experimental results demonstrate that \toolname{} fixes 326 out of 835 bugs in Defects4J, 24 out of 79 vulnerabilities in VUL4J, and achieves the best results on the multi-hunk subsets of SEC-bench and PatchEval.} Specifically, the \coordinator{}-\proposer{} framework enables \toolname{} to address more complex bugs, including 27 multi-file bugs and 62 multi-method bugs. Among the {}{95} unique bugs fixed by \toolname{}, {}{46} are multi-hunk, highlighting the agent's strong capability in multi-hunk repair. This emphasizes the importance of focusing on multi-hunk scenarios and suggests avenues for enhancing repair strategies in future research.
    
\end{itemize}

\section{Background and Motivation}

\subsection{LLM-based APR}
LLM-based APR~\cite{yang2025survey,xu2025aligning,li2025hybrid,farzandway2025automated,renzullo2025automated,yang2025revisiting,zhang2026effectiveness,zhang2026compass} builds on earlier pre-trained repair models~\cite{mashhadi2021applying,jiang2021cure,zhang2022coditt5}, which typically formulate repair as Neural Machine Translation (NMT~\cite{stahlberg2020neural}) or Masked Language Modeling (MLM~\cite{wang2019bert}). Although previous studies~\cite{zhang2023survey,xia2023automated} have shown the effectiveness of these paradigms, they still suffer from limitations such as dependence on large amounts of training data and limited generalization in real-world scenarios~\cite{zhang2024systematic}.
Compared with earlier pre-trained approaches, LLMs provide stronger repair capability with little or no task-specific training~\cite{zhang2024systematic}, as well as a broader range of available models, such as GPT-3~\cite{brown2020language} and GPT-4~\cite{achiam2023gpt}. Existing LLM-based APR methods can be broadly divided into training-free and training-based methods.

\textbf{Training-free APR.}
Training-free methods mainly include prompt-based and agent-based approaches. Prompt-based methods~\cite{cao2025study}, such as zero-shot~\cite{sobania2023analysis}, few-shot~\cite{nashid2023retrieval}, and Chain-of-Thought prompting~\cite{wei2022chain}, are popular for their simplicity. For example, ChatRepair~\cite{xia2024automated} interleaves patch generation with feedback, while ThinkRepair~\cite{yin2024thinkrepair} enhances repair through Chain-of-Thought prompting and iterative fixing. Agent-based methods~\cite{bouzenia2024repairagent,pabba2025semagent,yu2025patchagent,liu2024marscode,rondon2025evaluating} further equip LLMs with external tools and iterative workflows for planning, execution, and refinement. For instance, RepairAgent~\cite{bouzenia2024repairagent} combines ChatGPT with a state machine to autonomously plan and refine repair steps.

\textbf{Training-based APR.}
Training-based methods mainly include fine-tuning and reinforcement learning. Fine-tuning-based methods improve repair performance by adapting pre-trained models to repair tasks~\cite{drain2021generating,yuan2022circle,hao2023enhancing,silva2025repairllama}. Reinforcement learning has also shown promise in APR, as demonstrated by methods such as SWE-RL~\cite{wei2025swe} and Repair-R1~\cite{hu2025repair}. Despite these advances, most existing LLM-based APR methods still focus on relatively simple bugs or treat repair as a single-step generation problem.

\subsection{Multi-hunk Bug Repair}
Most previous APR studies focus on relatively simple bugs with concentrated fault locations. For example, ChatRepair~\cite{xia2024automated} and ThinkRepair~\cite{yin2024thinkrepair} mainly evaluate single-line, single-hunk, and single-method bugs. Recent studies~\cite{xia2023automated,xiang2024far,campos2025empirical,huang2025dynafix} also narrow their scope to function-level issues.
By contrast, research specifically targeting multi-hunk repair remains limited. On the one hand, multi-hunk bugs often span multiple methods or files, making repair substantially more difficult. On the other hand, many existing methods~\cite{xia2024agentless,liu2024marscode,xia2024automated} still adopt the same one-shot strategy used for simpler bugs, without explicitly modeling hunk dependencies or repair order. ITER~\cite{ye2024iter} is among the first studies to highlight the distinction between single-hunk and multi-hunk repair and proposes iterative refinement for such bugs. More recent work investigates context granularity and repair strategies for multi-hunk scenarios~\cite{nashid2025characterizing,gharibi2025multimend,nashid2025beyond}.
{Among these efforts, PReMM~\cite{xie2025premm} is closest to our setting: it decomposes multi-method bugs through dependency-based faulty-method clustering and repairs them with a fixed divide-and-conquer strategy, whereas \toolname{} treats repair as dynamic hunk-level coordination over an evolving partial patch.}
{}
{Together, these studies suggest that multi-hunk repair requires not only broader context or iterative retry, but also explicit coordination of the repair trajectory across dependent hunks. In this paper, we address this need through a \coordinator{}-\proposer{} framework.}

\subsection{Motivation}
\label{sec:motivation}

{}
{}
{}
{}

{Figure~\ref{fig:motivation_example} presents a motivation example from Defects4J Mockito\_17. The user expects a mock to be both serializable and configured with extra interfaces. However, the original implementation realizes \texttt{serializable()} by replacing the extra-interface configuration with \texttt{Serializable.class}, so the two requirements cannot hold at the same time and the serialization test fails. Fixing this bug is not a local edit: one hunk must record the serialization intent, while another hunk must later consume this intent when composing the final mock interfaces. This example motivates our design from two aspects.}

{\textbf{Generation-Recognition Asymmetry.} Our design is inspired by the Generation-Recognition Asymmetry phenomenon~\cite{peyrichou2026generation,fan2024oracle}: recognizing a correct solution from candidates can be easier than generating it directly. In Mockito\_17, GPT-3.5 fails to generate a correct patch after five self-correction attempts. However, in a diagnostic comparison, when we place the failed patches together with the developer patch and ask GPT-3.5 to choose from these candidates, GPT-3.5 can identify the correct patch. This gap suggests that LLMs are less reliable at judging an isolated self-generated patch than at comparing alternatives. It also reflects cognitive fixedness, where repeated self-correction tends to explore similar repair directions. The above observations motivate the propose-then-select design of \toolname{}.}

{\textbf{Repair-Order Scheduling.} The same example also motivates explicit hunk scheduling. The \texttt{MockSettingsImpl} hunk is a state producer: it changes \texttt{serializable()} from modifying \texttt{extraInterfaces} to recording an intent flag. The \texttt{MockUtil} hunk is a state consumer: it reads this flag and appends \texttt{Serializable.class} while preserving existing interfaces. Repairing the producer first gives the consumer a clear invariant to enforce, motivating a \coordinator{} that schedules hunks rather than repairing all locations independently.}

\begin{figure}[htbp]
\begin{tcolorbox}[colback=lightgray,colframe=deepblue,boxrule=0.6pt,left=2pt,right=2pt,top=2pt,bottom=2pt]
{\scriptsize\ttfamily
\setlength{\fboxsep}{1.2pt}
\textcolor{deepblue}{\textbf{Defects4J Mockito\_17: serializable() + extraInterfaces(List.class)}}\\[2pt]
\diffhunk{H1: MockSettingsImpl.java}{// producer of serialization intent}\\
\diffdel{- serializable(): return extraInterfaces(Serializable.class)}\\
\diffadd{+ private boolean serializable;}\\
\diffadd{+ serializable(): serializable = true; return this;}\\
\diffadd{+ isSerializable(): return serializable;}\\[2pt]
\diffhunk{H2: MockUtil.java}{// consumer when composing interfaces}\\
\diffctx{\ \ interfaces = settings.getExtraInterfaces();}\\
\diffadd{+ if settings.isSerializable():}\\
\diffadd{+\ \ \ \ ancillaryTypes = append(interfaces, Serializable.class)}\\
\diffadd{+ else:}\\
\diffadd{+\ \ \ \ ancillaryTypes = interfaces or empty;}
}
\end{tcolorbox}
  \caption{{Motivation example from Defects4J Mockito\_17.}}
  \label{fig:motivation_example}
\end{figure}

{Overall, this example shows that multi-hunk repair requires both selecting correct edits from multiple candidates and coordinating dependent hunks in a meaningful order. A one-shot LLM repair must solve these two problems simultaneously, which makes the repair unstable. \toolname{} addresses this difficulty by separating candidate proposal, candidate selection, and hunk scheduling, so that the repair intent established by one hunk can guide the edit of another hunk.}

\section{Approach}
\subsection{Problem Statement}

Suppose $\mathcal{D} = \{(B_i, H_i, P_i)\}_{i=1}^{|\mathcal{D}|}$ is a defect dataset containing $|\mathcal{D}|$ bugs. Each bug $B_i$ is a multi-hunk bug composed of $|H_i|$ hunks, where $H_i = [h_1, \dots, h_n]$ and each hunk $h_j$ represents a contiguous block of erroneous code. The corresponding fix is a patch $P_i = [p_1, \dots, p_n]$, where $p_j$ denotes the correction for hunk $h_j$. The multi-hunk repair task can thus be formalized as follows:
\find{
\begin{definition}[Multi-hunk Repair Task]
\label{def:multi_hunk_repair}
Given a buggy program with $n$ hunks $H_i = [h_1, \dots, h_n]$, the goal is to generate a complete patch $P_i = [p_1, \dots, p_n]$ such that each $p_j$ correctly fixes $h_j$. This is modeled as an autoregressive sequence generation task:
\begin{equation}
\pi_\gamma(P_i \mid B_i, H_i) = \prod_{j=1}^{n} \pi_\gamma(p_j \mid h_1, \dots, h_j; p_1, \dots, p_{j-1}; B_i)
\end{equation}
where $\gamma$ are parameters of the repair model.
\end{definition}
}

{} {Based on this, \toolname{} introduces a set of \proposer{}s to repair individual hunks and generate hunk-level patches, and a \coordinator{} to select the most promising candidate patch under the current repair context, as well as to orchestrate the repair order of the hunks.} Thus, the \coordinator{}-\proposer{} repair framework can be formalized as:

\find{
\begin{definition}[\coordinator{}-\proposer{} Repair Framework]
\label{def:coordinator_proposer_repair}

The \coordinator{}-\proposer{} repair process is characterized by three sequential and iterative steps:

\smallskip
\noindent\textbf{(1) Hunk Selection by \coordinator{}:}  
The \coordinator{} selects the next hunk $h_{\text{next}}$ to repair based on previously repaired hunks and patches, modeling the repair order as a conditional policy:
\begin{equation}
h_{\text{next}} = h_j \quad \text{where} \quad j = \arg\max_{j \in \mathcal{U}} \, \Pr(h_j \mid H_i, P_{<j}, B_i; \theta_c)
\end{equation}
where $\mathcal{U}$ is the set of unpatched hunks, $H_i$ is the set of all buggy hunks, $P_{<j}$ denotes patches generated for prior hunks, and $\theta_c$ are parameters of the \coordinator{}'s policy.

\smallskip
\noindent\textbf{(2) Patch Proposal by \proposer{}s:}  
Given $h_{\text{next}}$, $K$ \proposer{}s independently generate candidate patches $\{c_k\}_{k=1}^K$, forming the candidate set $C = \{c_k\}_{k=1}^K$. Each patch is sampled from a proposal distribution:
\begin{equation}
c_k \sim p_\phi(c \mid h_{\text{next}}, B_i), \quad k = 1, \dots, K
\end{equation}
where $\phi$ represents the parameters of the \proposer{} models.

\smallskip
\noindent\textbf{(3) Patch Selection by \coordinator{}:}  
{}
{The \coordinator{} evaluates the candidate set and selects $c_{\text{opt}}$ by maximizing the estimated probability of correctness, where ``optimal'' denotes the candidate judged most promising under the current hunk context and partial repair state:}
\begin{equation}
c_{\text{opt}} = \arg\max_{c \in C} \, \Pr(\text{valid}(c) \mid c, h_{\text{next}}, B_i; \theta_c)
\end{equation}
{}
{$\text{valid}(c)$ denotes that candidate $c$ is syntactically correct and can successfully compile.}

\end{definition}
}

\subsection{Overview}
\begin{figure*}[htbp]
  \centering
  \includegraphics[width=1.0\linewidth]{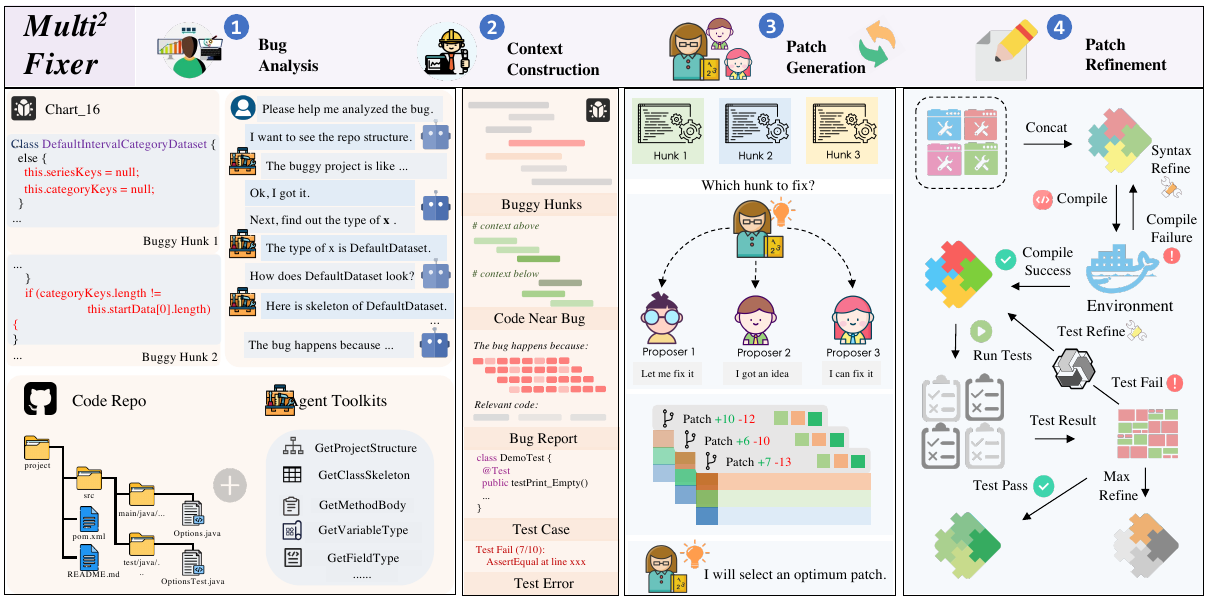}
  \caption{Overview of \toolname{}.} 
  \label{fig:overview}
\end{figure*}

Figure~\ref{fig:overview} illustrates the comprehensive workflow of \toolname{}, which is meticulously structured into four distinct yet interrelated phases: bug analysis, repair context construction, patch generation, and patch refinement. In the bug analysis phase, \toolname{} employs a tool-augmented bug analysis agent, \buganalyzer{}, to locate and analyze the root cause and relevant code of the bug, ultimately generating a bug report. In the repair context construction phase, \toolname{} collects five types of information as context for the repair model: the buggy code, surrounding code near the bug, the bug report generated in Phase 1, the failing test cases, and the test error report. In the patch generation phase, \toolname{} employs a \coordinator{}-\proposer{} architecture to perform heuristic patch search for multi-hunk bugs. {} {Specifically, \proposer{}s are a group of agents responsible for generating hunk-level candidate patches, while the \coordinator{} handles hunk scheduling and retains the candidate judged most promising under the current repair context.} Ultimately, \coordinator{} completes the repair of all hunks and assembles them into a complete patch. In the patch refinement phase, \toolname{} performs two-stage refinement on the patches generated in the previous stage. First, it compiles the patch; if compilation fails, it uses the compiler error messages to prompt the model for syntax refinement until the patch compiles successfully. Then, it applies the compiled patch to the buggy project, runs the relevant test cases, and iteratively refines the patch based on the test failure reports. In the following section, we provide a detailed description of each phase.

\subsection{Workflow of \toolname{}}

\subsubsection{Bug Analysis.} When facing a bug, \toolname{} first analyzes it by employing a dedicated tool-augmented agent called \buganalyzer{}, which leverages code-related tools to understand the issue. Specifically, \buganalyzer{} navigates the buggy codebase, gradually narrows down the bug's scope, and ultimately identifies its root cause.
To achieve this, we simulate the way human developers debug issues, modeling the process within an IDE environment (e.g., PyCharm). Initially, a developer examines the code corresponding to the error stack trace to gain insight into the failure. However, this part of code often represents only the tip of the iceberg, while the actual bug may reside in dependent or calling code. As a result, developers use their judgment to inspect suspicious variables, method calls, and other relevant elements in the buggy context.
Modern IDEs provide powerful features such as navigation, code folding, and expansion, enabling developers to efficiently explore class and method skeletons to locate the source of bugs. To emulate these capabilities, we equip \buganalyzer{} with a suite of tools that can inspect variable and field types, retrieve class skeletons, and obtain the project structure. These tools allow \buganalyzer{} to simulate IDE-like navigation and exploration through tool calls.
Once localization reaches the method level, developers typically examine the method’s implementation in detail to understand the bug. To support fine-grained analysis, we implement a method code viewing tool. Furthermore, recognizing that human short-term memory is limited, similar to the context window of LLMs, we design a context summarization tool that enables \buganalyzer{} to autonomously compress and manage its working context.
After multiple rounds of reasoning and tool invocations, \buganalyzer{} ultimately outputs a comprehensive bug report, including the root cause of the bug, the relevant code, and a preliminary solution for fixing the bug. All tools used by \buganalyzer{} are defined in Table~\ref{tab:tools}, we use JavaParser to perform static analysis on the buggy projects and implement these tools based on the analysis.
\begin{table}[t]
  \centering
  \caption{Bug analysis tools for \buganalyzer{}.}
  \label{tab:tools}
  \scriptsize
  \setlength{\tabcolsep}{3pt}
  \renewcommand{\arraystretch}{2.0}
  \newcommand{\apicell}[1]{\raisebox{0.3ex}{\parbox[c]{\linewidth}{\raggedright #1}}}
  \newcommand{\desccell}[1]{\raisebox{0.3ex}{\parbox[c]{\linewidth}{#1}}}
  \begin{tabular}{>{\raggedright\arraybackslash}m{0.30\linewidth}|>{\raggedright\arraybackslash}m{0.62\linewidth}}
      \hline
      \textbf{API {}{Name}} & {\textbf{Description and Output}} \\
      \hline
      \apicell{GetProjectStructure ( )} & {\desccell{\textbf{Description:} Get the repository file structure.\\[-0.2ex]\textbf{Output:} The repository file structure.}} \\
      \hline
      \apicell{GetImportOfFile (\texttt{file})} & {\desccell{\textbf{Description:} Get the imports of a given file.\\[-0.2ex]\textbf{Output:} The imports of the file.}}   \\
      \hline
      \apicell{GetClassSkeleton (\texttt{class})} & {\desccell{\textbf{Description:} Get the skeleton of a class.\\[-0.2ex]\textbf{Output:} Class skeleton.}} \\
      \hline
      \apicell{GetFieldType (\texttt{class}, \texttt{field})} & {\desccell{\textbf{Description:} Get the type of a class variable.\\[-0.2ex]\textbf{Output:} Fully qualified name of the class variable's type.}} \\
      \hline
      \apicell{GetVariableType (\texttt{class},\\ \texttt{method}, \texttt{variable})} & {\desccell{\textbf{Description:} Get the type of a local variable.\\[-0.2ex]\textbf{Output:} Fully qualified name of the local variable's type.}} \\
      \hline
      \apicell{GetMethodBody\\(\texttt{class}, \texttt{method})} & {\desccell{\textbf{Description:} Get the code of a class method.\\[-0.2ex]\textbf{Output:} The code of the class method.}} \\
      \hline
      \apicell{SummarizeContext\\(\texttt{summarizedContext})} & {\desccell{\textbf{Description:} Condense excessive context.\\[-0.2ex]\textbf{Output:} The compressed context given by the agent.}}\\
      \hline
      \apicell{Exit ( )} & {\desccell{\textbf{Description:} Exit if the agent has finished analyzing the bug.\\[-0.2ex]\textbf{Output:} N/A}} \\
      \hline
  \end{tabular}
  \normalsize
\end{table}

\subsubsection{Repair Context Construction.} After completing the analysis of the bug, \toolname{} begins constructing the repair context. Specifically, \toolname{} considers five types of contextual information, including: (1) \textbf{Buggy hunks.} Since the buggy code directly reflects the exact location of the error and represents the most critical information for repair, we first incorporate it into the repair context in the form of a hunk list. At the same time, we include necessary contextual information such as the class and method in which the hunk is located. (2) \textbf{Code near the buggy hunks.} In addition to the buggy hunks, the code near the buggy hunk is also highly important. This surrounding code often provides dependency-related information required by the buggy hunk, such as variable declarations, method definitions, and control flow context. Therefore, we also include this nearby code in the repair context. Specifically, \toolname{} employs three different levels of granularity when extending the context: line-level, method-level, and class-level context, to capture relevant semantic information at varying scopes. Line-level context includes specific lines (e.g., 10, 20, 50) of code immediately before and after the buggy hunk. Method-level context provides the entire method containing the bug. Class-level context encompasses the full definition of the buggy class.
(3) \textbf{Test code.} To keep the repair context concise, we only include the code of the failing test cases. (4) \textbf{Test failure report.} The failure report contains abundant information about the cause of the error, which can greatly help the model understand and fix the bug. To reduce the context length, we only extract the first five lines of each failure report. (5) \textbf{Bug report.} Finally, we include the bug report generated by \buganalyzer{} in Phase 1, which leverages repository-level context and provides understanding of the bug and repair guidance, further enhancing the model’s ability to generate accurate fixes. Together, all five types of contexts form a comprehensive and multi-granular representation that enables precise and informed patch generation.

\subsubsection{Patch Generation.} After constructing the context, \toolname{} proceeds to patch generation through a \coordinator{}-\proposer{} framework. In this framework, the \coordinator{} is responsible for scheduling the repair order of hunks and selecting the most promising patch, while multiple \proposer{}s are responsible for generating candidate edits for the selected hunk.

{\textbf{Hunk Scheduling.}} {} {We design the hunk scheduling mechanism based on the characteristics of multi-hunk bugs.} By analyzing multi-hunk bugs and their corresponding developer patches in Defects4J, we observe three representative repair patterns. First, in \textbf{Symmetric Fixing}, multiple hunks share similar bug patterns, so a repair strategy found for one hunk can often be reused or adapted for others. For example, in Chart\_19, null checks need to be inserted for two different objects; although the target variables differ, the repair structure is highly similar. Second, in \textbf{Stepwise Fixing}, some hunks have clear prerequisite relations, such as introducing a field before initializing it or defining a helper function before calling it. For instance, in Cli\_39, a file-stream creation function must be implemented before the faulty call site can be corrected. Third, in \textbf{Change Propagation}, a code change in one location requires coordinated updates in other locations to preserve consistency. For example, in Closure\_64, modifying the parameter list of the \texttt{to\_source} method requires corresponding changes at its call sites.

{}
{These observations suggest that multi-hunk repair should not be treated as a one-shot generation problem. Instead, the repair process should explicitly account for inter-hunk dependencies and repair order. To this end, the \coordinator{} first uses static-analysis tools together with the bug report to construct a hunk-dependency graph:}
\begin{equation}
{G=(H,E), \quad e_{ab}=(h_a,h_b,r_{ab},w_{ab})}
\end{equation}
{where $H$ is the set of buggy hunks, $r_{ab}$ denotes the relation type between two hunks, and $w_{ab}$ denotes the relation confidence. The first three relation types correspond to the repair patterns discussed above, while \textit{failure relevance} links hunks to the bug report or failure message. During repair, the \coordinator{} selects the next hunk according to this dependency graph and the current repair state:}
\begin{equation}
{h_{\mathrm{next}}=\mathcal{C}(G,S_t,\mathcal{U}_t)}
\end{equation}
{where $\mathcal{U}_t$ is the set of unrepaired hunks. After a hunk-level patch is accepted, the repair state and dependency graph are updated, and the \coordinator{} uses the updated graph to choose the next hunk.}

{\textbf{Patch Proposal and Selection.}} {} {After selecting the next hunk, the \coordinator{} sends the target hunk and the current repair state to $K$ \proposer{}s, which independently generate a candidate set $C_t=\{c_1,\ldots,c_K\}$. The \coordinator{} first normalizes these candidates and groups similar candidates into clusters $\mathcal{P}_t=\{P_1,\ldots,P_m\}$. For each candidate $c$, we define its cluster confidence as $\kappa(c)=|P(c)|/K$, where $P(c)$ is the cluster containing $c$; candidates supported by larger clusters receive higher confidence. We model the \coordinator{}'s selection as estimating three normalized criteria in $[0,1]$: $s_{\mathrm{ctx}}(c)$ for consistency with the bug report and local code context, $s_{\mathrm{state}}(c)$ for compatibility with already accepted patches, and $s_{\mathrm{rel}}(c)$ for consistency with the hunk-dependency graph and usefulness for subsequent related hunks. The final selection objective is:}
\begin{equation}
{S(c)=\kappa(c)+s_{\mathrm{ctx}}(c)+s_{\mathrm{state}}(c)+s_{\mathrm{rel}}(c), \quad
c^*=\arg\max_{c\in C_t}S(c)}
\end{equation}
{Before accepting $c^*$, the \coordinator{} applies a code-format fallback check. If the selected candidate violates basic structural constraints, such as inconsistent indentation, unbalanced brackets, invalid function-signature replacement, or copying surrounding context outside the target hunk, \toolname{} falls back to the highest-scoring structurally valid candidate. If no generated replacement is structurally safe for a replacement hunk, the original hunk is preserved to avoid introducing a compilation failure; for insertion hunks, the highest-scoring non-empty candidate is used. Once a candidate is accepted, the partial patch is updated and the \coordinator{} proceeds to the next repair step.} {} {We define a repair trajectory as the ordered sequence of \coordinator{} decisions, selected hunks, accepted candidate patches, and any revisits made during one repair round. If this trajectory later proves inconsistent, the \coordinator{} may revisit a previously repaired hunk and revise its earlier decision.}

{}
{This mechanism operationalizes the generation-recognition motivation in Section~\ref{sec:motivation}: \proposer{}s explore alternatives, while the \coordinator{} compares, scores, and filters them under the current repair state. The process iterates until all hunks are repaired or the maximum number of iterations is reached.}

{} {}

\subsubsection{Patch Refinement.} {} {After Phase 3 assembles a complete patch, we use a dedicated refining model to perform two-stage patch refinement in Phase 4; here, a valid patch refers to a patch that is syntactically well-formed and can compile successfully.} First, we perform syntax refinement on the patch if it fails to compile. During patch generation, models may introduce extraneous context or produce syntactic errors such as incorrect indentation, missing, or mismatched parentheses. To address these issues, we employ the refining model to carefully inspect and correct such errors through iterative feedback, ensuring syntactic validity.
After iterative syntax refinement and compilation, if the patch successfully compiles, we proceed to test refinement. In this stage, the same refining model is further leveraged to interpret test failure reports and iteratively revise the patch when test cases fail. This process continues until the patch passes all test cases or reaches the maximum number of iterations.

\section{Experimental Setup}
\subsection{Research Questions}
We evaluate \toolname{} on the following research questions:

\noindent\textbf{RQ1:} How does \toolname{} compare against the state-of-the-art APR techniques?

\noindent\textbf{RQ2:} How does \toolname{} perform with different base models?

\noindent\textbf{RQ3:} How does each component contribute to the performance of \toolname{}?

\noindent\textbf{RQ4:} How does \toolname{} perform when extended to vulnerability repair tasks?

\noindent\textbf{RQ5:} How does the cost of \toolname{} compare to baselines?

\subsection{Datasets}
We evaluate \toolname{} on {}{Defects4J~\cite{just2014defects4j} and three vulnerability benchmarks: VUL4J~\cite{bui2022vul4j}, SEC-bench~\cite{lee2025secbench}, and PatchEval~\cite{wei2025patcheval}}. Defects4J~\cite{just2014defects4j} is a collection of bugs from real Java open-source projects, including 395 bugs from Defects4J-v1.2 and 440 bugs from Defects4J-v2. VUL4J~\cite{bui2022vul4j} consists of 79 reproducible, real-world Java
vulnerabilities corresponding to 51 open-source projects. {SEC-bench~\cite{lee2025secbench} contains 200 C++ security tasks, from which we select 65 multi-hunk cases. PatchEval~\cite{wei2025patcheval} contains 1,000 real-world vulnerabilities in JavaScript, Go, and Python, from which we select 120 multi-hunk cases with runnable Docker environments.} Specifically, to better investigate \toolname{}'s repair performance across different types of bugs, we follow previous works~\cite{xie2025premm,xia2024automated} to identify six types of bugs based on the location and scope of the faulty code, including single-line, single-hunk, single-method, multi-method, single-file, and multi-file bugs.

\subsection{Baselines}
We compare \toolname{} against {}{eleven} state-of-the-art APR baselines, including ChatRepair~\cite{xia2024automated}, ThinkRepair~\cite{yin2024thinkrepair}, RepairAgent~\cite{bouzenia2024repairagent}, GiantRepair~\cite{li2025hybrid}, ContrastRepair~\cite{kong2025contrastrepair}, RepairLLaMA~\cite{silva2025repairllama}, ITER~\cite{ye2024iter}, MultiMend~\cite{gharibi2025multimend}, BIRCH~\cite{nashid2025characterizing}, {PReMM~\cite{xie2025premm},} and NTR~\cite{ntr}. These baselines cover prompt-based, agent-based, fine-tuning-based, and multi-hunk APR methods. Unless otherwise noted, baseline results are taken directly from the original papers. If a paper does not report a particular result, we leave the corresponding table entry blank.

\subsection{Evaluation Metrics}
We consider two widely used metrics~\cite{zhao2024enhancingautomatedprogramrepair,xin2024practicalusefulautomatedprogram,yang2024revisitingunnaturalnessautomatedprogram} to evaluate the effectiveness of both \toolname{} and baselines, and the quality of the generated patches. The definitions of the metrics are listed as follows. {}
{Correct Fix (CF) is defined as the number of bugs for which the generated patch passes all available tests and is manually checked for semantic correctness with respect to the developer patch. In this manual check, semantic correctness means that the generated patch preserves the intended program behavior captured by the developer patch, rather than merely passing the available tests.}
{}
{Plausible Fix (PF) is defined as the number of bugs for which the generated patch is valid and passes all available tests, but no manual semantic-equivalence check is applied.}

\subsection{Implementation Details}
We implement \toolname{} using both API-based models (e.g., GPT-3.5) and open-source models (e.g., Qwen2.5-72B). In our experiments, we use oracle function-level buggy locations provided by the benchmark setting as input to the repair framework. To control cost, we limit the number of tool invocations in the bug analysis phase to at most 20. During multi-hunk repair, we treat the repair of each individual hunk as one step and set the maximum number of hunk-repair steps to 20 per repair round. We dynamically assign different temperatures to the \proposer{}s, with the temperatures uniformly distributed in the range [0, 1]. For example, when using three \proposer{}s, their temperatures are set to 0, 0.5, and 1.0, respectively. In the patch refinement phase, we perform syntax refinement and test refinement, each with a maximum of three iterations. For each bug, we run five repair rounds.

\noindent\textbf{Patch Size.}
We define patch size as the number of complete patch candidates validated for each bug. Let $R$ denote the number of repair rounds, and let $I_{\text{syn}}$ and $I_{\text{test}}$ denote the maximum numbers of syntax-refinement and test-refinement iterations, respectively. The patch size is bounded by:
\begin{equation}
B_{\text{patch}} \le R \times (1 + I_{\text{syn}} + I_{\text{test}})
\end{equation}
where the term 1 denotes the initial patch generated before refinement. In our experiments, we set $R=5$, $I_{\text{syn}}=3$, and $I_{\text{test}}=3$, yielding $B_{\text{patch}} \le 5 \times (1+3+3) = 35$. Here, only complete patches submitted to compilation and testing are counted toward patch size, while intermediate hunk-level edits are not counted. Notably, the patch size used in our experiments is relatively small compared with that used in many existing studies.
We implement \toolname{} based on the PyTorch and Transformers frameworks. All experiments are conducted on two NVIDIA Tesla V100 GPUs on a single Ubuntu 20.04 server.

\section{Evaluation and Results}
\subsection{RQ1: Comparison with State-of-the-Arts}

\textbf{Experimental Design.} 
In RQ1, we aim to evaluate the performance of \toolname{}. We select {}{eleven} state-of-the-art APR methods as baselines, including both LLM-based (e.g., ChatRepair) and agent-based (e.g., RepairAgent) approaches, and compare their repair performance with \toolname{} on a total of 835 bugs in Defects4J. We use GPT-3.5 as the base model to facilitate comparison with prior APR studies.

\noindent\textbf{Overall Performance.} 
{}
{Table~\ref{tab:overall_performance} presents the comparison results on Defects4J. Since prior baselines are evaluated under different scopes, we use the Scope column to distinguish full-benchmark and subset results; cross-scope comparisons are reported only as contextual evidence. On the full Defects4J-v1.2+v2 benchmark with 835 bugs, \toolname{} fixes 326 bugs and generates 412 plausible patches. Among baselines evaluated on the same 835-bug scope, \toolname{} outperforms PReMM, the strongest full-scope baseline, by 19 correct fixes and 37 plausible fixes.}
{For multi-hunk repair, the advantage of \toolname{} is more visible on complex bug categories. \toolname{} fixes 62 multi-method bugs and 27 multi-file bugs, exceeding PReMM (45 multi-method and 15 multi-file bugs), MultiMend (17 multi-method and 9 multi-file bugs), and RepairAgent (7 multi-method and 3 multi-file bugs). In contrast, on simpler categories such as SL, SH, and SM, the performance gap is smaller, suggesting that \toolname{}'s main benefit comes from coordinating repairs across multiple hunks, methods, or files.}

\begin{table*}[htbp]
\centering
\caption{Overall performance of \toolname{}, Single Line (SL), Single Hunk (SH), Single Method(SM), Multiple Methods (MM), Single File (SF), Multiple files (MF), Correct Fix (CF), Plausible Fix (PF). We report only the results explicitly provided in the original baseline papers; missing entries are denoted by \textbf{-}. {Scope denotes the number or subset of Defects4J bugs evaluated in the original study; in the Scope column, MH indicates the multi-hunk subset.}}
\label{tab:overall_performance}

\resizebox{\linewidth}{!}{
\begin{tabular}{l c c | c c c c c c c c | c c c c c c c c|c c}
\toprule
\multirow{2}{*}{\textbf{Method}} & \multirow{2}{*}{{\textbf{Scope}}} & \multicolumn{1}{c|}{\multirow{2}{*}{\textbf{Patch Size}}} & \multicolumn{8}{c|}{\textbf{D4J-v1.2}} & \multicolumn{8}{c|}{\textbf{D4J-v2.0}} & \multicolumn{2}{c}{\textbf{Total}} \\
& & & \textbf{SL} & \textbf{SH} & \textbf{SM} & \textbf{MM} & \textbf{SF} & \textbf{MF} & \textbf{CF} & \textbf{PF}  &\textbf{SL} & \textbf{SH} & \textbf{SM} & \textbf{MM} & \textbf{SF} & \textbf{MF}& \textbf{CF} & \textbf{PF} & \textbf{CF} & \textbf{PF}\\
\midrule
ITER (ICSE, 2024) & {476} & 1000 & 36 & 44 & 52 & 7 & 54 & 5 & 59 & 89 & 9 & 15 & 16 & 3 & 16 & 3 & 36 & 19 & 78 & 125\\
ThinkRepair (ISSTA, 2024)  & {483 SM} & $\leq$125 & 52 & 78 & 98 & 0 & 98 & 0 & 98 &-
& 47 & 81 & 107  & 0 & 107 & 0 & 107 &- & 205 & -\\
ChatRepair (ISSTA, 2024) & {337} & $\leq$500            &   57 & 79 & 114 & 0 &  114 & 0 & 114   &  - &  48 & 48 & 48 & 0 & 48 & 0 & 48 & - & 162 & -\\
RepairAgent (ICSE, 2025) & {835} &  $\approx$117  & 52 & 67 & 86 & 4 & 88 & 2 & 90 & 96  &    48 & 61 & 71 & 3 & 73 & 1  & 74 & 90 & 164 & 186\\
RepairLLaMA (TSE, 2025) & {483} & 10 & - & - & - & - & - & - & - & - & 90 & 139 & 142 & 2 & 144 & 0 & 144 & 195 & 144 & 195 \\
GiantRepair (TOSEM, 2025) & {483 SM} & $\leq$200 & 28 & 48 & 52 & 1 & 53 & 0 & 53 & - & 28 & 52 & 51 & 2 & 53 & 0 & 53 & - & 106 & - \\ 
ContrastRepair (TOSEM, 2025) & {337} & $\leq$160 & 60 & 99 & 101 & 2 & 103 & 0 & 103 & - & 40 & 40 & 40 & 0 & 40 & 0 & 40 & - & 143 & 201\\
MultiMend (arXiv, 2025) & {835} & $\geq$100 & 47 & 61 & 67 & 12 & 74 & 5 & 79 & 127 & 37 & 63 & 65 & 5 & 66 & 4 & 70 &132 & 149 & 259\\
BIRCH (ASE, 2025) & {372 MH} & 1 & - & - & - & - & - & - & - & 83 &  - & - & - & - & - & - & - &  50 & - & 133\\
NTR (ICSE, 2025) & {D4J-v1.2} & 100 & - & - & - & - & - & - & 139 & 177 &  - & - & - & - & - & - & - &  - & 139 & 177\\
{PReMM (OOPSLA, 2025)} & {835} & {15}  &    {53} & {70} & {121} & {26} & {140} & {7}  & {147} & {184} &   {52} & {129} & {141} & {19} & {152} & {8} & {160} & {191} & {307} & {375}\\
\toolname{} (Ours)  & {835} & $\leq$35 &  49 & 74 & 120 & 28 & 141 & 7 & 148 & 182 & 56 & 98 & 144 & 34 & 158 & 20  & 178 & 230 & 326 & 412\\
\bottomrule
\end{tabular}}
\end{table*}

\begin{figure}[htbp]
\centering
    \subfigure[Venn on Defects4J-v1.2]{
        \includegraphics[width=0.47\columnwidth]{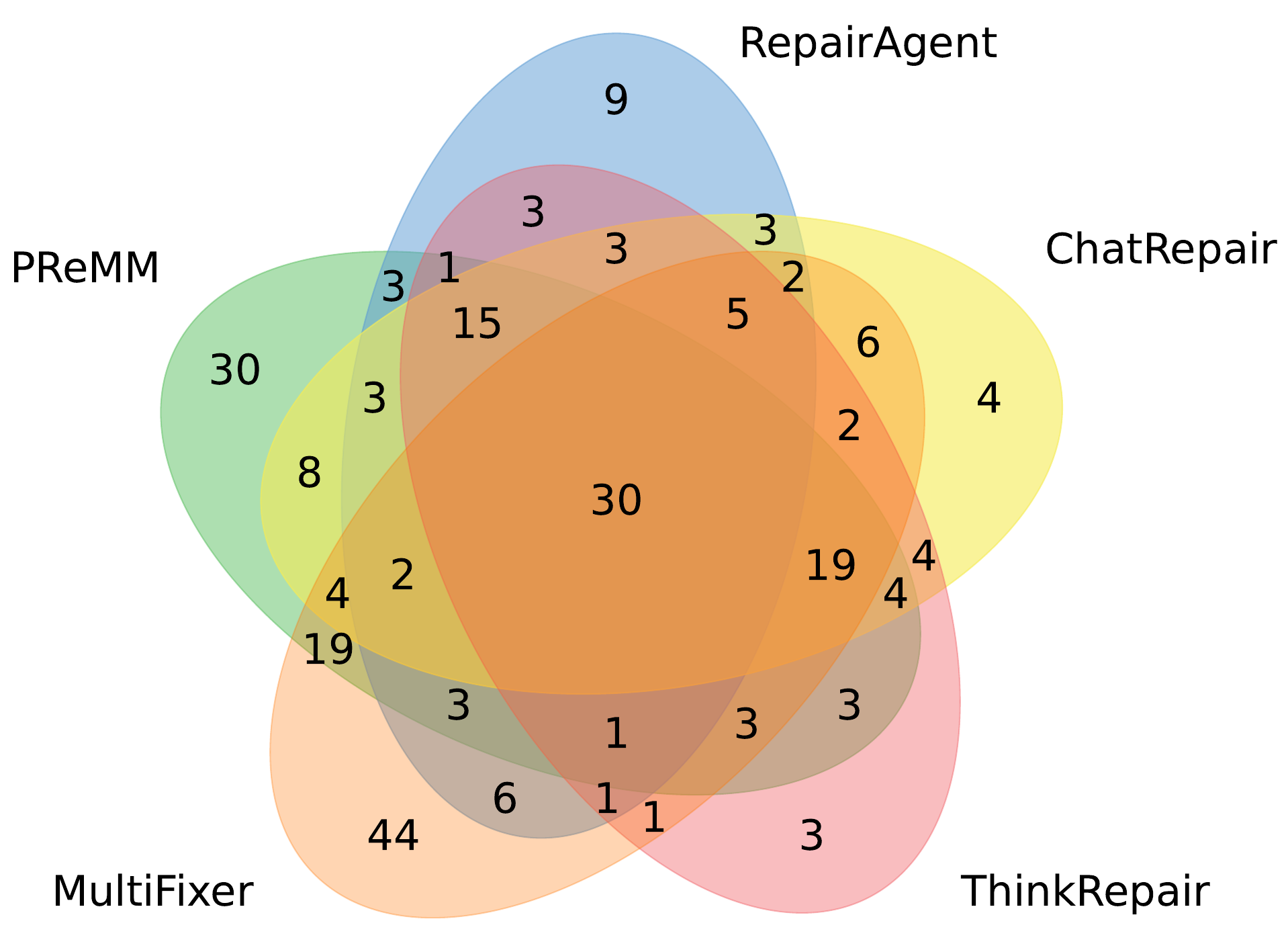}
        \label{venn_d4j12}
    }
    \subfigure[Venn on Defects4J-v2] {
        \includegraphics[width=0.47\columnwidth]{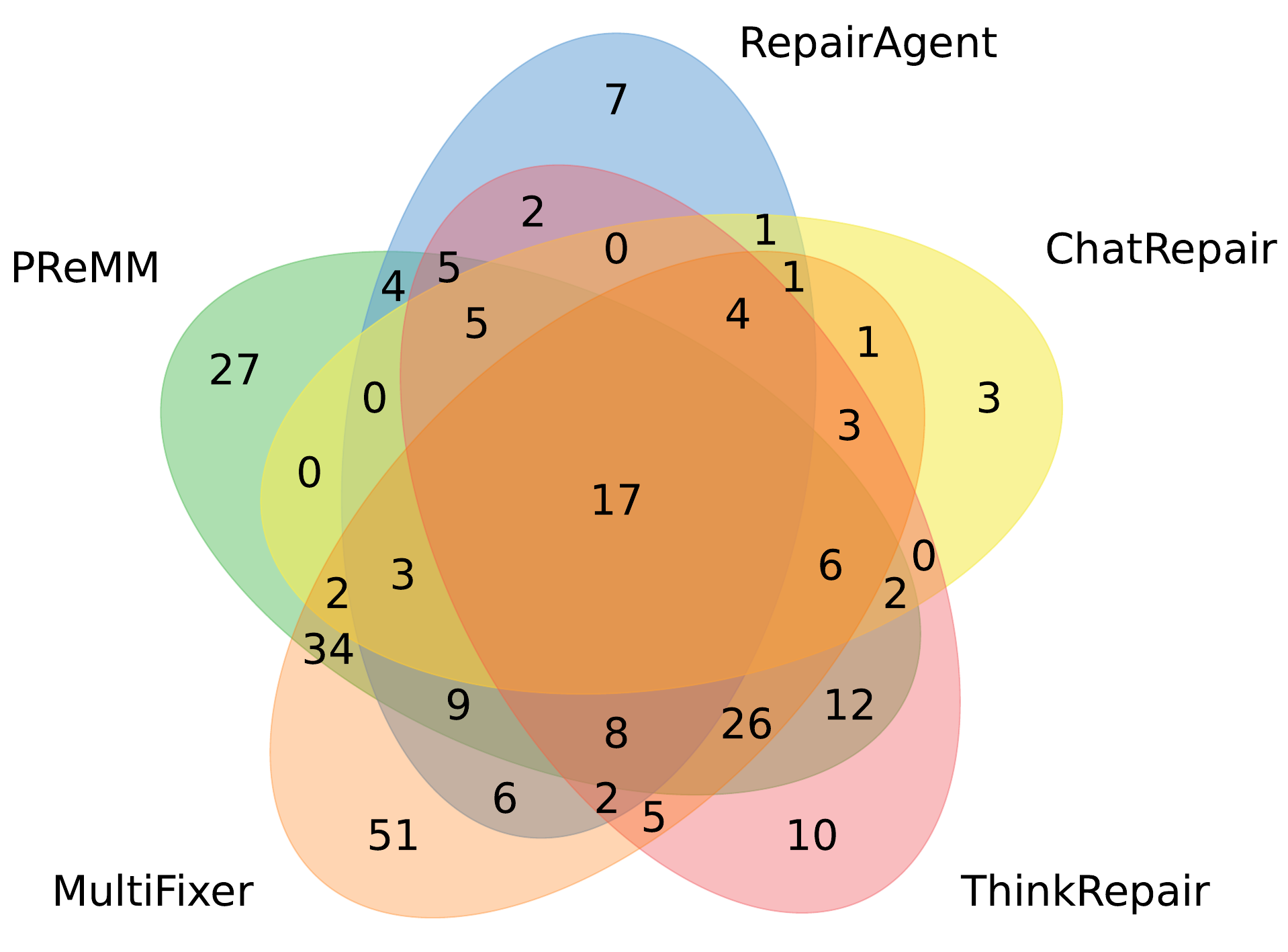}
        \label{venn_d4j2}
    }
    \caption{Bugfix Venn Diagram on Defects4J (\toolname{}, ThinkRepair, ChatRepair, RepairAgent, {}{PReMM})}
    \label{d4j_venn}
\end{figure}

\noindent\textbf{Overlap Analysis.} We select the top four baselines with the best repair results for the overlap analysis.
Figure \ref{d4j_venn} shows the Venn diagram of the bugs fixed by RepairAgent~\cite{bouzenia2024repairagent}, ChatRepair~\cite{xia2024automated}, ThinkRepair~\cite{yin2024thinkrepair}, {}{PReMM~\cite{xie2025premm}} and \toolname{} on Defects4J-v1.2 and Defects4J-v2. Figure \ref{d4j_venn} shows that \toolname{} {}{fixes additional unique bugs, including} {}{44} and {}{51} unique bugs on DefectsJ-v1.2 and v2, respectively, compared to the other 4 baselines. We further observe that among the {}{95} unique bugs fixed by \toolname{}, {}{46} are multi-hunk,  highlighting \toolname{}'s strong capability in multi-hunk repair. Additionally, Table~\ref{tab:baseline_detail} compares the distribution of fixed bugs across different methods from the perspective of buggy projects. \toolname{} shows balanced repair performance across all projects and {}{performs well} on projects such as JxPath, JacksonDatabind, and Math.

\begin{table*}[htbp]
  \centering
\caption{Distribution of bugs fixed by \toolname{} across projects in Defects4J. Core is short for JacksonCore, Xml is short for JacksonXml, Databind is short for JacksonDatabind, Collect is short for Collections.}
\label{tab:baseline_detail}
\tabcolsep=2pt
    \resizebox{0.9\linewidth}{!}{\begin{tabular}{c|ccccccccccccccccc|c}
    \toprule
     \toolname{} & Closure & Chart & Lang  & Math & Mockito & Time & Cli & Codec&  Collect & Compress & Csv & Gson & Core & Databind & Xml & JxPath & Jsoup & Total\\
    \midrule

    \# Bugs & 174 & 26 & 64 & 106 & 38 & 26 & 39 & 18 & 4 & 47 & 16 & 18 & 26 & 112 & 6 & 22 & 93 & 835 \\
    \midrule
    Plausible & 55 & 15 & 36 & 54 & 20 & 11 & 22 & 11 & 2 & 28 & 12 & 9 & 14 & 62 & 3 & 9 & 49 & 412 \\
    Correct & 41  & 15 & 29 & 45 & 18 & 7 & 17 & 10 & 2 & 19 & 8 & 9 & 11 & 48 & 3 & 8 & 36 & 326 \\
    \midrule
    {PReMM} & {37} & {22} & {36} & {41} & {9} & {4} & {17} & {8} & {0} & {22} & {10} & {7} & {13} & {44} & {1} & {1} & {35} & {307} \\
    ThinkRepair & 34 & 11 & 19 & 27 & 6 & 4 & 9 & 10 & 0 & 16 & 8 & 5 & 7 & 17 & 2 & 2 & 28 & 205 \\
    RepairAgent & 27 & 11 & 17 & 29 & 6 & 2 & 8 & 9 & 1 & 10 & 6 & 3 & 5 & 11 & 1 & 0 & 18 & 164 \\
    ChatRepair & 37 & 15 & 21 & 32 & 6 & 3 & 5 & 8 & 0 & 2 & 3 & 3 & 3 & 9 & 1 & 0 & 14 & 162 \\
    MultiMend & 27 & 8 & 16 & 23 & 6 & 1 & 11 & 6 & 1 & 8 & 4 & 3 & 2 & 12 & 2 & 3 & 16 & 149\\
    RepairLLaMA & 21 & 9 & 13 & 24 & 4 & 3 & 6 & 3 & 1 & 10 & 4 & 5 & 3 & 15 & 0 & 0 & 23 & 144\\
    ContrastRepair & 32 & 12 & 19 & 30 & 8 & 2 & 4 & 5 & 0 & 2 & 3 & 1 & 3 & 7 & 1 & 0 & 14 & 143\\
    NTR & 40 & 14 & 29 & 39 & 12 & 5 & 0 & 0 & 0 & 0 & 0 & 0 & 0 & 0 & 0 & 0 & 0 &  139 \\
    GiantRepair & 18 & 7 & 5 & 20 & 3 & 1 & 4 & 6 & 0 & 7 & 5 & 3 & 6 & 10 & 1 & 0 & 10 & 106\\
    ITER & 18 & 10 & 10 & 20 & 0 & 2 & 6 & 3 & 0 & 4 & 2 & 0 & 3 & 0 & 0 & 0 & 0 & 78\\

    \bottomrule
    \end{tabular}}
    \end{table*}

\subsection{RQ2: Performance of Different Base Models}
\textbf{Experimental Design.} 
In RQ1, we use GPT-3.5 as the base model to facilitate comparison with prior APR studies on Defects4J. To further investigate the impact of different base models on \toolname{}, in RQ2 we select several currently mainstream models beyond GPT-3.5, including Claude-3.5-Sonnet, DeepSeek-V3.2-Exp, Qwen2.5-Max, and the open-source model Qwen2.5-72B-Instruct, as base models for \toolname{}, and evaluate their repair performance on Defects4J, respectively.

\noindent\textbf{Results and Analysis.} As shown in Table~\ref{tab:different_base_models}, all five base models demonstrate remarkable repair capability on Defects4J when integrated with \toolname{}. Specifically, Claude-3.5-Sonnet significantly outperforms the other models, generating 514 plausible fixes and correctly repairing 420 of them, achieving a repair rate of 50.29\%. Especially on various types of multi-hunk bugs, Claude demonstrates balanced performance due to its deep understanding of syntactic and semantic relationships across different code locations. Qwen2.5-Max ranks second overall, repairing 71 fewer bugs than Claude-3.5-Sonnet. DeepSeek-V3.2-Exp performs well on single-hunk bugs, ranking second among the five models, but shows poor performance on multi-hunk bugs, repairing only 55 MM bugs and 20 MF bugs, placing last among the five models. This not only highlights the significant differences in structural characteristics and repair difficulty between single-hunk and multi-hunk bugs, but also reflects DeepSeek-V3.2-Exp's inability to maintain balanced and comprehensive performance across different types of bugs. Notably, the smaller-scaled model Qwen2.5-72B-Instruct also achieves promising repair results on Defects4J, fixing only 15 fewer bugs than GPT-3.5. This demonstrates that \toolname{} can effectively adapt to LLMs of different sizes and maintain strong performance across diverse base models.

\begin{table*}[htbp]
\centering
\caption{Performance of \toolname{} using different base models.}
\label{tab:different_base_models}

\resizebox{0.8\linewidth}{!}{
\begin{tabular}{l | c c c c c c c c | c c c c c c c c | c c}
\toprule
\multirow{2}{*}{\textbf{Model}} 
    & \multicolumn{8}{c|}{\textbf{D4J-v1.2}} 
    & \multicolumn{8}{c|}{\textbf{D4J-v2.0}} 
    & \multicolumn{2}{c}{\textbf{Total}} \\
& \textbf{SL} & \textbf{SH} & \textbf{SM} & \textbf{MM} & \textbf{SF} & \textbf{MF} & \textbf{CF} & \textbf{PF} 
& \textbf{SL} & \textbf{SH} & \textbf{SM} & \textbf{MM} & \textbf{SF} & \textbf{MF} & \textbf{CF} & \textbf{PF} 
& \textbf{CF} & \textbf{PF} \\
\midrule
GPT-3.5           &  49 & 74 & 120 & 28 & 141 & 7 & 148 & 182 & 56 & 98 & 144 & 34 & 158 & 20  & 178 & 230 & 326 & 412 \\
Claude-3.5 & 59 & 98 & 157 & 37 & 183 & 11 & 194 & 232 & 68 & 115 & 178 & 48 & 199 & 27 & 226 & 282 &  420 & 514 \\
DeepSeek-V3.2  & 57 & 96 & 137 & 26 & 158 & 5 & 163 & 198 & 62 & 101 & 147 & 29 & 161 & 15 & 223 & 238 & 339 & 421\\
Qwen2.5-Max  & 51 & 82 & 133 & 35 & 159 & 9 & 168 & 203 & 56 & 96 & 145 & 36 & 157 & 24 & 181 & 234 & 349 & 437 \\
Qwen2.5-72B  & 47 & 69 & 116 & 26 & 135 & 7 & 142 & 175 & 54 & 93 & 139 & 30 & 152 & 17 & 169 & 224  & 311 & 399\\
\bottomrule

\end{tabular}
}
\end{table*}

\subsection{RQ3: Ablation Study}
\noindent\textbf{Experimental Design.} 
In RQ1 and RQ2, we have compared \toolname{} with existing methods and discussed the differences in repair effectiveness among various base models on the full Defects4J benchmark. In the subsequent ablation study, we focus more on the effectiveness of each component of \toolname{} in the context of multi-hunk bug repair. Specifically, we analyze and discuss the effectiveness of the four components used by \toolname{}, using a total of 372 multi-hunk bugs from Defects4J. We systematically remove each component from \toolname{} one at a time, and compare the repair performance of \toolname{} before and after the removal of each individual component.

\noindent\textbf{RQ3.1: Effectiveness of Bug Analysis.} In RQ3.1, we first discuss the impact of bug analysis on repair effectiveness. As shown in Table~\ref{tab:ablation_study_bug_analysis}, all five models show improved repair performance after incorporating bug analysis. Among them, GPT-3.5 and Claude-Sonnet-3.5 show the most significant improvement, fixing 18 and 17 additional multi-hunk bugs, respectively, and both achieve 24 more plausible fixes. In contrast, bug analysis has a weaker effect on DeepSeek-V3.2-Exp, merely enabling the model to fix 9 additional bugs, which is consistent with DeepSeek-V3.2-Exp's overall poorer performance in fixing multi-hunk bugs. Moreover, we find that bug analysis not only enables the model to successfully fix bugs that were previously completely unreparable (i.e., no plausible fix), but also helps the model achieve a more comprehensive understanding of bugs that already had plausible fixes but were not fully resolved. This leads to more complete and refined repairs. Such improvements are particularly evident in scenarios involving edge cases and complex conditional logic.
\begin{table}[hthp]
\centering
\caption{Ablation study on bug analysis.}
\resizebox{\linewidth}{!}{
\begin{tabular}{lccccc}
\toprule
 & \textbf{GPT-3.5} & \textbf{Claude-3.5} & \textbf{DeepSeek-V3.2} & \textbf{Qwen2.5-Max} & \textbf{Qwen2.5-72B} \\ 
\midrule
w/o Bug Analysis           &  71/100    & 112/138 & 63/85 & 96/117 &  68/95      \\
w/ Bug Analysis        &   89/124    & 129/162   & 72/94    & 106/135  &  82/108     \\
\bottomrule
\end{tabular}}
\label{tab:ablation_study_bug_analysis}
\end{table}

\noindent\textbf{RQ3.2: Effectiveness of Repair Context.} In RQ3.2, we compare the impact of three different context granularities on the repair outcomes. As shown in Table~\ref{tab:ablation_study_context}, we find that the performance of line-level and method-level contexts is comparable, while class-level context yields poorer repair results. After detailed analysis, we identify two reasons for the poorer performance of class-level context. First, some classes contain numerous functions unrelated to the repair task, causing the repair context to exceed the model's context window and directly leading to repair failure. Second, the redundant context introduces irrelevant information that interferes with the model’s judgment, causing it to incorrectly associate the bug with unrelated code and resulting in erroneous fixes. For line-level context, we experiment with different lengths ranging from 5 to 50 in increments of 5. By comparing the results, we find that a context length of 20 achieves the best trade-off between performance and cost. This result is comparable to that of method-level context, because a properly sized line-level context effectively preserves the relevant information within the method. In contrast, although method-level context achieves good repair performance, it sometimes leads to information redundancy and increased cost when the buggy method is excessively long (e.g., more than 300 lines). Thus, we conclude that line-level context offers greater flexibility and strikes a better balance between cost and performance, making it more suitable for \toolname{}.
\begin{table}[hthp]
\centering
\caption{Ablation study on context granularity.}
\resizebox{\linewidth}{!}{
\begin{tabular}{lccccc}
\toprule
\textbf{Context Granularity} & \textbf{GPT-3.5} & \textbf{Claude-3.5} & \textbf{DeepSeek-V3.2} & \textbf{Qwen2.5-Max} & \textbf{Qwen2.5-72B} \\ 
\midrule
Class-Level Context      & 79/100     & 115/143  & 64/82 & 97/124 &  72/88  \\ 
Method-Level Context        & 86/122   & 132/167  & 70/91 & 110/141 &  78/101         \\
Line-Level Context   &   89/124     &  129/162 & 72/94 & 106/135 & 82/108   \\

\bottomrule
\end{tabular}}
\label{tab:ablation_study_context}
\end{table}

\noindent\textbf{RQ3.3: Effectiveness of the \coordinator{}-\proposer{} Framework.} 
{}
{}
{}
{In RQ3.3, we compare \toolname{} with three scheduling variants. To isolate the effect of hunk scheduling, Sequential Scheduling, Random Scheduling, and \toolname{} all use the same number of \proposer{}s (proposer=3), so the comparison focuses on the scheduling strategy. \textbf{(1) w/o Scheduling} removes explicit hunk scheduling and generates patches for all hunks at once as a whole-patch repair. \textbf{(2) Sequential Scheduling} repairs hunks one by one according to their order in the developer patch. \textbf{(3) Random Scheduling} shuffles the sequential hunk order using seed=42 and then repairs hunks one by one according to the shuffled order. As shown in Table~\ref{tab:ablation_study_coordinator_proposer}, \toolname{} consistently outperforms all three variants. Compared with w/o Scheduling, \toolname{} fixes 18--56 more bugs across the five base models. Compared with Sequential Scheduling and Random Scheduling, \toolname{} further fixes 2--13 and 3--16 more bugs, respectively. These results show that the improvement comes not only from iterative hunk-level repair, but also from dynamic scheduling by the \coordinator{}.}

\noindent{\textbf{Sensitivity of \proposer{} Number.} We conduct a sensitivity experiment on the number of \proposer{}s by sampling 100 multi-hunk bugs from Defects4J. When increasing the number of \proposer{}s from 3 to 5, patch selection becomes less stable across three runs and shows a 3--10\% performance drop. This suggests that a larger candidate pool can introduce noisy or conflicting patches that outweigh useful diversity.}
\begin{table}[hthp]
\centering
\caption{Ablation study on the \coordinator{}-\proposer{} framework.}
\resizebox{\linewidth}{!}{
\begin{tabular}{lccccc}
\toprule
{\textbf{Scheduling Strategy}} & \textbf{GPT-3.5} & \textbf{Claude-3.5} & \textbf{DeepSeek-V3.2} & \textbf{Qwen2.5-Max} & \textbf{Qwen2.5-72B} \\ 
\midrule
{w/o Scheduling}      & {52/78}  & {73/94} & {54/75} & {61/83} & {47/72}     \\
{}{Sequential Scheduling (proposer=3)}      & {}{81/108}  & {}{116/147} & {}{70/91} & {}{96/121} & {}{76/97}     \\
{Random Scheduling (proposer=3)}      & {79/104}  & {113/143} & {69/90} & {96/123} & {78/100}     \\
{\coordinator{}-\proposer{} (proposer=3)}        &   89/124    & 129/162   & 72/94    & 106/135  &  82/108     \\
\bottomrule
\end{tabular}}
\label{tab:ablation_study_coordinator_proposer}
\end{table}

\noindent\textbf{RQ3.4: Effectiveness of Patch Refinement.} In RQ3.4, we discuss the impact of syntax refinement and test refinement on repair effectiveness separately. First, regarding syntax refinement, we find that the model often correctly understands the bug and generates generally correct patches, but may produce syntactic flaws (e.g., extra braces or inclusion of context code). Upon being prompted about such issues, the model is generally able to quickly fix syntax errors. Therefore, as shown in Table~\ref{tab:ablation_study_refinement}, we find that syntax refinement significantly improves repair effectiveness, with the five models fixing 19 to 38 additional bugs after applying syntax refinement.  In contrast, test refinement is more challenging. It requires prompting the model with failing test information to correct the patch, which in turn demands that the model understand the test report and make substantial modifications to the patch. Therefore, the effectiveness of test refinement is somewhat limited, yet it still helps the models fix an additional 17 to 24 bugs. By combining the two refinement strategies, the model can proceed from easy to difficult, first correcting simple syntax errors, and then further refining the semantics of the patch.
\begin{table}[hthp]
\centering
\caption{Ablation study on patch refinement.}
\resizebox{\linewidth}{!}{
\begin{tabular}{lccccc}
\toprule
\textbf{Refine Strategy} & \textbf{GPT-3.5} & \textbf{Claude-3.5} & \textbf{DeepSeek-V3.2} & \textbf{Qwen2.5-Max} & \textbf{Qwen2.5-72B} \\ 
\midrule
w/o Refinement           &   40/49  & 72/77 & 33/43 & 50/58 &   38/45        \\

\quad+ Syntax Refinement        &   67/87 
& 105 /134 
& 52/66
& 88/112
&    65/84     \\

\quad\quad+ Test Refinement      
&   89/124 
& 129 /162 
& 72/94
& 106/135
&    82/108     \\ 

\bottomrule
\end{tabular}}
\label{tab:ablation_study_refinement}
\end{table}

\subsection{RQ4: Performance of \toolname{} on Vulnerability Repair}
\noindent\textbf{Experimental Design.} 
{}
{In the previous RQs, we have evaluated \toolname{} on the test-driven bug dataset Defects4J. To assess its generalizability to vulnerability repair, RQ4 uses three vulnerability benchmarks. First, we evaluate on VUL4J~\cite{bui2022vul4j}, including 35 single-hunk and 44 multi-hunk vulnerabilities. On VUL4J, we compare \toolname{} with FSV~\cite{effective}, NTR~\cite{ntr}, VRPILOT~\cite{vrpilot}, APR4Vul~\cite{bui2024apr4vul}, and ChatRepair~\cite{xia2024automated}. Second, to further examine multi-hunk vulnerability repair beyond Java, we evaluate the multi-hunk subsets of SEC-bench~\cite{lee2025secbench} (C++, N=65) and PatchEval~\cite{wei2025patcheval} (JavaScript, Go, and Python, N=120), and compare \toolname{} with ChatRepair~\cite{xia2024automated}, PReMM~\cite{xie2025premm}, and MultiMend~\cite{gharibi2025multimend} under the same GPT-3.5 setting.}

\begin{table*}[hthp]
\centering
\footnotesize
\caption{Comparison results between \toolname{}-Vul and existing baselines on VUL4J. {$^{*}$ denotes \toolname{} without oracle fault localization.}}
\resizebox{1.0\linewidth}{!}{
\begin{tabular}{c|cccccccc}
\toprule
\textbf{}        & \textbf{\toolname{}} & {\textbf{\toolname{}$^{*}$}} & \textbf{FSV-Codex~\cite{effective}} & \textbf{FSV-finetuned~\cite{effective}} & \textbf{NTR~\cite{ntr}} & \textbf{VRPILOT~\cite{vrpilot}} & \textbf{APR4Vul~\cite{bui2024apr4vul}} & \textbf{ChatRepair~\cite{xia2024automated}}\\ 
\midrule
\textbf{Single-Hunk}   & 19/35     & {19/35}     & 11/35      & 9 /35             & 14/35  & 14/35  & 16/35 & 14/35   \\

\textbf{Multi-Hunk}   & 5/44     & {3/44}     & 0/44      & 0/44              & 0/44  & 0/44  & 0/44 & 1/44 \\
\textbf{Total}   & 24 (30.37\%)     & {22 (27.85\%)}     & 10.9 (13.79\%)      & 9 (11.39\%)             & 14 (17.72\%)  & 14 (17.72\%)  & 16 (20.25\%) & 15 (18.98\%)  \\
\bottomrule 
\end{tabular}}
\label{tab:vul}
\end{table*}

\begin{table}[t]
\centering
\footnotesize
\caption{{Repair results on the multi-hunk subsets of SEC-bench and PatchEval.}}
\label{tab:additional_vul}
\resizebox{\linewidth}{!}{\begin{tabular}{clccccc}
\toprule
{\textbf{Dataset}}
& {\textbf{Method}}
& {\textbf{SM}}
& {\textbf{MM}}
& {\textbf{SF}}
& {\textbf{MF}}
& {\textbf{Total}} \\
\midrule

\multirow{4}{*}{  {\shortstack{\textbf{SEC-bench}\\\textbf{(N=65)}}}}
& {ChatRepair}
& {0}
& {9}
& {5}
& {4}
& {9} \\

& {PReMM}
& {0}
& {2}
& {1}
& {1}
& {2} \\

& {MultiMend}
& {0}
& {5}
& {4}
& {1}
& {5} \\

& {\toolname{}}
& {0}
& {11}
& {8}
& {3}
& \textbf{{11}} \\

\midrule

\multirow{4}{*}{  {\shortstack{\textbf{PatchEval}\\\textbf{(N=120)}}}}
& {ChatRepair}
& {1}
& {5}
& {5}
& {1}
& {6} \\

& {PReMM}
& {1}
& {16}
& {15}
& {2}
& {17} \\

& {MultiMend}
& {1}
& {7}
& {6}
& {2}
& {8} \\

& {\toolname{}}
& {3}
& {16}
& {16}
& {3}
& \textbf{{19}} \\

\bottomrule
\end{tabular}}
\end{table}

\noindent{} 
{}
{}
{}
{}
\noindent{\textbf{Results on VUL4J.} Table~\ref{tab:vul} shows the VUL4J results. With GPT-3.5 as the base model, \toolname{} fixes 24 vulnerabilities, outperforming the second-best baseline APR4Vul by 8 fixes. Among these fixes, 5 are multi-hunk vulnerabilities, while the strongest baseline fixes at most 1 multi-hunk vulnerability. The \toolname{}$^{*}$ result further shows that removing oracle fault localization preserves single-hunk performance (19/35) and still fixes 3 multi-hunk vulnerabilities, indicating that \toolname{} remains effective in realistic end-to-end vulnerability repair.}
\par\noindent{\textbf{Results on SEC-bench and PatchEval.} Table~\ref{tab:additional_vul} reports the results on the multi-hunk subsets of SEC-bench and PatchEval. On SEC-bench, \toolname{} fixes 11 vulnerabilities, outperforming ChatRepair with 9 fixes, MultiMend with 5 fixes, and PReMM with 2 fixes. On PatchEval, \toolname{} fixes 19 vulnerabilities, also outperforming PReMM with 17 fixes, MultiMend with 8 fixes, and ChatRepair with 6 fixes. These results show that \toolname{}'s advantage is consistent on non-Java vulnerability benchmarks and is especially visible on multi-hunk cases, where explicit hunk scheduling and hunk-level patch selection help coordinate edits across different vulnerable locations.}

\subsection{RQ5: Cost Analysis of \toolname{}}
\noindent\textbf{Experimental Design.} 
In addition to evaluating \toolname{}'s repair capability, we further assess its practicality by computing the average resource consumption per bug, including the number of tokens used, time cost, and monetary cost, and compare these metrics with those of the baselines. 

\noindent\textbf{Results and Analysis.} As shown in Table~\ref{tab:cost_analysis}, we report the average cost of \toolname{} on single-hunk and multi-hunk bugs, respectively. Since multi-hunk bug repair relies on multiple \proposer{}s to generate patches independently, each \proposer{}'s conversation incurs additional cost, and the \coordinator{}'s scheduling also contributes to the overall expense. Therefore, the total cost of \toolname{} (MH) is significantly higher than that of \toolname{} (SH), approximately 2.5 times as much. However, compared to ChatRepair and RepairAgent, \toolname{} adopts a significantly smaller patch size (e.g., $\leq$35), which is less than one-third of that used in prior works, thereby achieving substantial performance advantages. Specifically, \toolname{} reduces repair time by 33\% and monetary cost by 53\% across all types of bugs.
\begin{table}[htbp]
    \centering
    \caption{Cost analysis between \toolname{}, ChatRepair and RepairAgent on Defects4J.}
    \resizebox{\linewidth}{!}{\begin{tabular}{lccccc}
    \toprule
    Method & Patch/Bug & Time/Bug & Token/Bug & Money/Bug & Charge/1k tokens\\
    \midrule
        ChatRepair (2024){}& $\leq$ 500 & $\leq$ 5h & 210,000 & \$0.42 & \$0.002\\

        RepairAgent (2024){} & 117  & 920s & 270,000 & \$0.54 & \$0.002 \\

        \toolname{} (SH) & $\leq35$ & 373s & 60,000& \$0.12& \$0.002\\
         \toolname{} (MH) & $\leq35$ & 929s & 150,000& \$0.3& \$0.002\\
          \toolname{} ({}{Avg}) & $\leq35$ & 620s & 100,000& \$0.2& \$0.002\\
 
    \bottomrule
    \end{tabular}
    }
    \label{tab:cost_analysis}
\end{table}

\section{{Discussion}}
\subsection{{Data Leakage}}
{Since Defects4J and VUL4J contain bugs whose fixing dates may precede the release of modern LLMs (e.g., GPT-3.5 in 2022), model memorization could contaminate repair results. We mitigate and analyze this risk as follows.}

\noindent{\textbf{(1) Same Model and Settings.} We evaluate all compared methods with the same base model (GPT-3.5) and comparable parameter settings, making the relative comparison reliable.}

\noindent{\textbf{(2) Multiple Repair Types.} We evaluate both bug repair and vulnerability repair benchmarks to reduce the risk that the conclusion is driven by memorization of one repair type.}

\noindent{\textbf{(3) Multiple Programming Languages.} Beyond the Java-only Defects4J and VUL4J benchmarks, we further evaluate SEC-bench (C++) and PatchEval (JavaScript, Go, and Python), showing that \toolname{}'s advantage is not limited to Java.}

\noindent{\textbf{(4) Time-Separated Analysis.} We conduct a time-separated analysis on 2022+ vulnerabilities from SEC-bench and PatchEval. As shown in Table~\ref{tab:time_separated_vul}, \toolname{} fixes 5 vulnerabilities from 2022, 1 from 2023, and 4 from 2024, matching PReMM in total and outperforming the other baselines. These cases provide a stricter check than the original pre-2022 Java benchmarks because their fixing dates are after the release of GPT-3.5. Moreover, \toolname{} fixes the largest number of 2024 vulnerabilities, providing additional evidence that its advantage is not solely explained by benchmark memorization.}
\begin{table}[t]
\centering
\scriptsize
\setlength{\tabcolsep}{2pt}
\caption{{Time-separated vulnerability repair results on selected vulnerabilities fixed in 2022 or later.}}
\resizebox{\linewidth}{!}{\begin{tabular}{@{}l|l@{}}
\toprule
{\textbf{Method}} & {\textbf{CVEs fixed after 2022 using GPT-3.5}} \\
\midrule
{ChatRepair} &
{\makecell[l]{CVE-2022-29188, njs.cve-2023-27727, CVE-2024-5823, gpac.cve-2024-50665}} \\
\midrule
{PReMM} &
{\makecell[l]{CVE-2022-24065, CVE-2022-29188, CVE-2022-31145, njs.cve-2022-31306, CVE-2022-35936,\\
CVE-2022-37109, CVE-2023-39660, CVE-2023-40029, CVE-2024-5823, CVE-2024-39330}} \\
\midrule
{MultiMend} &
{\makecell[l]{CVE-2022-21699, CVE-2022-29188, libdwarf.cve-2022-32200, libiec61850.cve-2023-27772,\\
CVE-2023-40029, CVE-2024-5823}} \\
\midrule
{\toolname{}} &
{\makecell[l]{CVE-2022-21699, CVE-2022-23857, CVE-2022-29188, CVE-2022-31145, CVE-2022-35949,\\
CVE-2023-40029, CVE-2024-3571, CVE-2024-24747, gpac.cve-2024-50665, CVE-2024-53900}} \\
\bottomrule
\end{tabular}}
\label{tab:time_separated_vul}
\end{table}

\subsection{{Trajectory Analysis of Hunk Scheduling}}
{To better demonstrate the role of the proposed hunk scheduling mechanism, we conduct a trajectory case study on PatchEval CVE-2024-3571. This vulnerability is a path traversal bug in LangChain's \texttt{LocalFileStore}: user-controlled keys are checked only by a regex before being joined with the storage root, so malicious keys can escape the intended directory without canonicalization and root-containment checks. \toolname{} schedules the three hunks as H2 $\rightarrow$ H3 $\rightarrow$ H1. H2 first canonicalizes \texttt{root\_path} into a resolved absolute path, establishing the security invariant for the storage root. H3 then resolves each key-derived path and checks whether it remains under the canonical root, enforcing this invariant at the use site. H1 is repaired last to add the necessary import/scaffold edits. This trajectory follows a security-driven repair logic: first establish the root-path invariant, then enforce the invariant at the key-to-path conversion point, and finally add the supporting code required by these edits. In comparison, ChatRepair fails to produce parseable hunk updates, while PReMM and MultiMend generate patches that fail the PatchEval PoC. This case shows that, compared with a general agentic generate-select-refine pipeline, \toolname{}'s advantage lies in hunk scheduling: it identifies the semantic role and dependency of each hunk and repairs them in an order that first establishes and then enforces the security invariant.}

\subsection{{Failure Analysis}}
{Although \toolname{} is effective for multi-hunk bug repair, it can still fail in specific high-complexity scenarios. Our manual inspection identifies two dominant patterns. \textbf{(1) Excessive-Hunk Looping.} For bugs with many hunks (10+), such as Jsoup\_87, the \coordinator{} may exhaust the 20-step budget by repeatedly revising already repaired hunks; among 16 bugs with 10+ hunks, 11 (68.75\%) exhibit this behavior. \textbf{(2) Complex Inter-Hunk Dependency.} For bugs with multi-directional or multi-level dependencies, the correct edit for one hunk may depend on decisions made in several other hunks. In a random sample of 20 failed cases, 7 (35\%) cases, such as Chart\_18, involve such dependencies. These failures suggest that the main bottleneck is often scheduling and coordination under high hunk complexity, rather than simply generating individual hunk patches.}

\section{Threats to Validity}

\textbf{Internal Validity}.
Internal validity concerns experimental biases affecting evaluation fairness and consistency. Because LLM-based repair is stochastic, output randomness may introduce noise. We mitigate this by running \toolname{} for five repair rounds per bug under identical settings and validating all complete patch candidates. This reduces sampling-induced instability while bounding the patch size at 35, balancing effectiveness and efficiency. All experiments run on the same machine in isolated environments to minimize uncontrolled factors.

\noindent\textbf{External Validity}.
External validity concerns whether our findings generalize beyond the evaluated benchmarks. Because Defects4J may not capture the diversity of real-world defects, {we additionally evaluate \toolname{} on VUL4J, SEC-bench, and PatchEval, covering Java, C++, JavaScript, Go, and Python vulnerability repair. The results indicate that \toolname{} generalizes beyond test-driven bug repair and remains effective across vulnerability benchmarks.}

\section{Conclusion}
In this paper, we propose \toolname{}, a \coordinator{}-\proposer{} based multi-agent framework for fixing multi-hunk bugs. By combining tool-augmented bug analysis, fine-grained repair context construction, hierarchical patch generation, and iterative refinement, \toolname{} can effectively address the challenges of repair order scheduling, hunk-level patch generation, and hunk-level patch selection. {}{Across Defects4J, VUL4J, SEC-bench, and PatchEval, \toolname{} fixes 326 Defects4J bugs (62 multi-method and 27 multi-file), 24/79 VUL4J vulnerabilities (22 without oracle fault localization), and 11/65 and 19/120 multi-hunk vulnerabilities on SEC-bench and PatchEval.}  Our work demonstrates the power of collaborative agents in fixing multi-hunk bugs.

\section*{Acknowledgments}
This research was supported in part by Natural Science Foundation of Jiangsu Province (BK20251458), and Fundamental Research Funds for the Central Universities (AE89991/463).

\section*{Data Availability Statement}
To facilitate reproducibility and further research, we release the full implementation of \toolname{}, including the source code, experiment configurations, and data processing pipeline. The project is available at \textit{{}{https://zenodo.org/records/21223018}}.

\bibliographystyle{ACM-Reference-Format}
\bibliography{reference}

\end{document}